\documentclass{article}
\usepackage{amsmath,epsfig}
\usepackage[preprint]{spconfa4}
\usepackage{multirow}
\usepackage{multicol}
\usepackage{threeparttable}
\usepackage{fancyhdr}

\copyrightnotice{978-1-7281-1331-9/20/\$31.00~\copyright~2020~IEEE~\hfill}

\let\OLDthebibliography\thebibliography
\renewcommand\thebibliography[1]{
  \OLDthebibliography{#1}
  \setlength{\parskip}{0pt}
  \setlength{\itemsep}{0pt plus 0.3ex}
}

\begin{document}\sloppy

\def\x{{\mathbf x}}
\def\L{{\cal L}}

%

\title{Universal Adversarial Perturbations Generative Network for Speaker Recognition}
%
\name{Jiguo Li$^{1,2,4}$, Xinfeng Zhang${^{4}}$, Chuanmin Jia${^{2}}$, Jizheng Xu$^3$, Li Zhang$^3$, Yue Wang$^3$, Siwei Ma$^{2\dagger}$, Wen Gao$^2$\thanks{The work was done when Jiguo Li interned in Bytedance.Inc. This work was supported by the National Natural Science Foundation of China and Royal Society (61961130392), National Natural Science Foundation of China (61632001), MoE-China Mobile Research Fund Project (MCM20180702), and High-performance Computing Platform of Peking University, which are gratefully acknowledged.}\thanks{$^\dagger$Siwei Ma~(swma@pku.edu.cn) is the corresponding author.}}
\address{$^1$Institute of Computer Technology, Chinese Academic of Sciences, jiguo.li@vipl.ict.ac.cn \\$^2$Peking University, \{cmjia, swma, wgao\}@pku.edu.cn \\$^3$Bytedance.Inc, \{xujizheng, lizhang.idm, wangyue.v\}@bytedance.com \\$^4$University of Chinese Academic of Sciences, xfzhang@ucas.ac.cn}

\maketitle

\begin{abstract}
  Attacking deep learning based biometric systems has drawn more and more attention
  with the wide deployment of fingerprint/face/speaker recognition systems, given the fact that the neural networks are vulnerable to the adversarial examples, which have been intentionally perturbed to remain almost imperceptible for human. 
  In this paper, we demonstrated the existence of the universal adversarial perturbations~(UAPs) for the speaker recognition systems. We proposed a generative network to learn the mapping from the low-dimensional normal distribution to the UAPs subspace, then synthesize the UAPs to perturbe any input signals to spoof the well-trained speaker recognition model with high probability.
  Experimental results on TIMIT and LibriSpeech datasets demonstrate the effectiveness of our model. 
  \end{abstract}
  \section{Introduction}
  \label{sec:intro}
  With the success of deep neural networks~(DNNs) since Krizhevsky~\textit{et al.}~\cite{krizhevsky2012imagenet} won the ImageNet challenge~\cite{russakovsky2015imagenet} in 2012, more and more deep-based models for biometric systems, such as fingerprint/face/speaker recognition, have been deployed in our daily life. However, these systems are facing the risk of being attacked since deep models are vulnerable to adversarial examples~\cite{goodfellow2014explaining}, which have been intentionally perturbed. Meanwhile, attacking the deep models and finding the weaknesses of the models can help us avoid the potential risk and design corresponding methods to defense against these attacks. 
  In these widely deployed biometric systems, previous works mainly focus on the vision-based systems, the audio-based systems, such as speaker recognition, have not been well-studied, although the speaker recognition systems have been widely deployed. 
  In this paper, we focus on the attack for speaker recognition models by generating the universal adversarial perturbations~(UAPs), which are independent of the input samples and can be applied to the whole dataset.
  \\
  \indent Before UAPs have been found by Moosavi-Dezfooli~\textit{et al.}~\cite{Moosavi2017Universal}, generating the adversarial examples and spoofing the well-trained deep models have become an emerging topic since Szegedy~\textit{et al.}~\cite{szegedy2013intriguing} found DNNs are vulnerable to the adversarial examples with intentional imperceptible perturbations. 
  Following~\cite{szegedy2013intriguing}, some other optimization methods, such as Adam~\cite{carlini2017towards}, Fast Gradient Sign Method~(FGSM)~\cite{goodfellow2014explaining} or the genetic algorithm~\cite{su2019one} are used to find the perturbations for the input image. 
  Recently, Moosavi-Dezfooli~\textit{et al.}~\cite{Moosavi2017Universal} demonstrated that there exists a universal and small perturbation that can spoof the well-trained DNN image classifier with high probability. Subsequently, Hayes~\textit{et al.}~\cite{hayes2018learning} crafted the UAPs by leveraging a generative network to synthesize the perturbation from the input noise which samples from the normal distribution, and improved the attack success rate as well as showed the transferability cross different models for the same dataset. Motivated by the existence of UAPs in the image classification, in this paper, we attempt to find the UAPs of the speaker recognition systems by designing a generative model~\cite{hayes2018learning}.
  \\
  \indent
  In addition to attacking the vision-based systems, the attack for speaker recognition systems has also been addressed for a long time. Before the DNNs have been used in the speaker recognition, the replay and synthesis attacks had been studied to avoid the risk in the voice verification systems~\cite{korshunov2016overview}. In recent years, with the wide deployment of DNN-based systems, attacking the DNN-based speaker recognition models has drawn more and more attention. Gong~\textit{et al.}~\cite{gong2017crafting} crafted the adversarial examples using FGSM to attack the well-trained speech verification model and showed the deep models are vulnerable to the adversarial attack. However, the evidence is missing on large-scale datasets~\cite{gong2017crafting}. Using the same optimization method, Kreuk~\textit{et al.}~\cite{kreuk2018fooling} presented white box attacks for text-dependent speaker verification on the deep end-to-end network on NTIMIT~\cite{jankowski1990ntimit} and YOHO~\cite{campbell1995testing}.
  \\
  \indent
  In this paper, we attempt to generate the UAPs by learning the mapping from the low-dimensional normal distribution to the universal perturbation subspace via a generative model, given the fact that the UAPs are not unique~\cite{Moosavi2017Universal}. We demonstrate the effectiveness of our proposed method by attacking the state-of-the-art speaker recognition model~\cite{ravanelli2018speaker} under non-targeted and targeted settings on TIMIT~\cite{garofolo1993darpa} and LibriSpeech~\cite{panayotov2015librispeech} datasets.
  Our contributions can be summarized as follows:
  \begin{itemize}
    \item We demonstrate the existence of the UAPs for the well-trained speaker recognition model, which are the potential risks for the widely deployed speaker recognition systems in our daily life.
    \item We can synthesize different UAPs efficiently by mapping the normal distribution into the UAPs subspace using the generative model. The experimental results show that our model can achieve an SER of 97.0 with an SNR of 49.87 and a PESQ of 3.00 in the non-targeted attack on TIMIT dataset, indicating the effectiveness of our proposed model.
    \item The ablation study for the UAPs shows that our proposed model can learn useful~\textit{universal patterns}, map the low-dimensional normal distribution into the UAPs subspace, and generate UAPs that perform much better than the random perturbations.
  \end{itemize}
  
  \section{Related Works}
  \label{sec:related_works}
  \subsection{UAPs Generation}
  The existence of UAPs have been demonstrated in many areas
  ~\cite{wu2019g, neekhara2019universal}
  , since Moosavi-Dezfooli~\textit{et al.}~\cite{Moosavi2017Universal} found the UAPs in the image classification. Here we mainly review some UAPs generation models on image classification and audio-based systems that are related to our work. Different from the iterative optimization method used in~\cite{Moosavi2017Universal}, Hayes~\textit{et al.}~\cite{hayes2018learning} crafted the UAPs by leveraging a generative network to synthesize the perturbation from the input noise which samples from the normal distribution, and improved the attack success rate as well as showed the transferability cross different models for the same dataset. In addition to the works in image classification, some works about UAPs generation for audio-based systems are also proposed recently. 
  Neekhara~\textit{et al.}~\cite{neekhara2019universal} iteratively searched the UAPs with minimal norm under the constraint of high attacking success rate, and only one UAP can be found in once optimization. 
  Our work is different from the above two works in two aspects: (1) our work focuses on the unexplored task, speaker recognition, to study the potential risk of the widely deployed authentication systems; (2) our generative attacker can synthesize different UAPs efficiently once trained, which has been demonstrated more effective than the iterative methods in image classification attacks~\cite{hayes2018learning}.
  \subsection{Speaker Recognition Attack}
  Attacking the speaker recognition models has drawn the researchers' attention because: (1) deep model attacking has become a hot topic in the machine learning community; (2) the speaker recognition/verification systems have been widely deployed in our daily life. Gong~\textit{et al.} crafted the adversarial examples iteratively to attack the speaker recognition model trained on a small dataset and demonstrated the existence of the adversarial example for the speaker recognition models. Subsequently, Kreuk~\textit{et al.}~\cite{kreuk2018fooling} attempted to fool the end-to-end speaker verification model which is trained on MFCC features by optimizing the perturbation using FGSM~\cite{goodfellow2014explaining}. However, these white-box attacks need gradients in the testing phase. In this paper, we proposed a semi-white attack model to learn the UAPs, which is more practical than the white box methods in the real scenario because: (1) our generative attacker needs no gradient in the testing stage; (2) the adversarial perturbations are universal and they can spoof the well-trained speaker recognition model with any input speeches.

  \begin{figure*}[t]
    \centering{\includegraphics[width=0.9\linewidth]{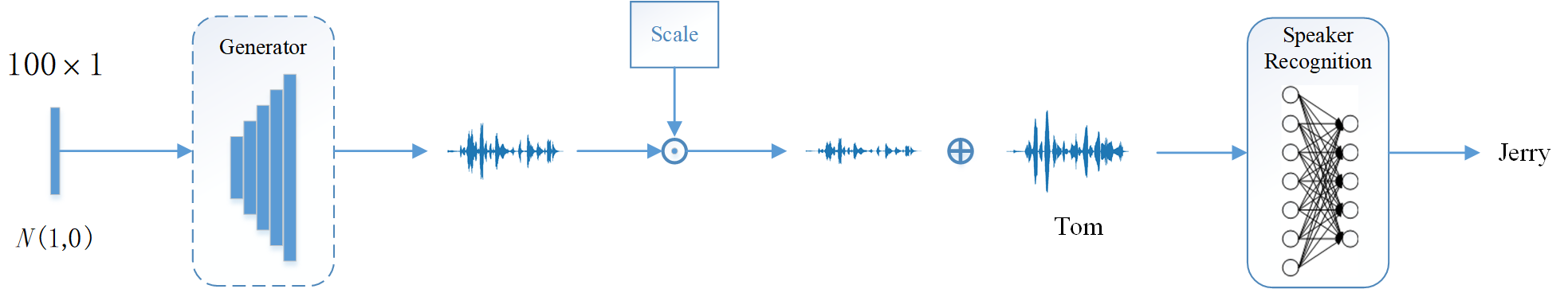}}
    \caption{The framework for our proposed universal adversarial perturbations generative network for speaker recognition.}
    \label{fig:proposed_model}
    \vspace{-5mm}
  \end{figure*}
  
  
  \section{Proposed Method}
  \label{sec:proposed_method}
  As illustrated in Fig.~\ref{fig:proposed_model}, our generative attacker aims to map the input noise, which is sampled from the low-dimensional normal distribution~$\mathcal{N}(0,1)$, into a UAP, and the following well-trained speaker recognition model is spoofed by the input adversarial example, which is perturbed by the generated UAP. Given a speech $s$ and its speaker label $y$, the non-targeted attack for the speaker recognition model with UAPs can be formulated as:
  \begin{align}
    \arg\min_{s'}{L(s, s+\delta)}&\quad\text{s.t.}\quad f(s+\delta)=y'\label{eq:non_tgt_attack}\\
    &\text{where}~\delta=G_{\theta}(z)~\text{and}~y'\neq y,\nonumber
  \end{align}
  where $G$ is the generative attack model to synthesize the UAP $\delta$ from the noise $z$, $y'$ is the prediction of the adversarial example $s'=s+\delta$, $f$ is a well-trained state-of-the-art speaker recognition model, $L$ is a distance function to measure the distortion between the raw signal and the adversarial example. For the targeted attack, we modify the constraint for $y'$ from $y'\neq y$ as $y'= y_t$, in which $y_t$ is the target class.
  \subsection{The Victim Model}
  We use the state-of-the-art speaker recognition model SincNet~\cite{ravanelli2018speaker} as our target victim model. SincNet achieved state-of-the-art performance on TIMIT~\cite{garofolo1993darpa} and LibriSpeech~\cite{panayotov2015librispeech} datasets by replacing the first convolution layer as the learnable band pass filters. Given the frequency band $[f_1, f_2]$, the learnable band pass filter can be described as:
  
  \begin{align}
    h[n,f_1,f_2] = 2f_2\text{sinc}(2\pi f_2 n)-2f_1\text{sinc}(2\pi f_1 n),\label{eq:band_pass_filter}
  \end{align}
  where $\text{sinc}(x)=\sin{x}/{x}$, and $f_1, f_2$ are the learnable parameters. By using the band pass filters rather than the convolution filters in the first layer, the model is more interpretable and achieves better results~\cite{ravanelli2018speaker}.
  
  \subsection{The Framework}
  Our model aims to spoof the well-trained speaker recognition model with UAPs. As illustrated in Fig.~\ref{fig:proposed_model}, the \textit{Generator} is a generative network with several upsampling blocks to synthesize the UAP from the input noise with 100 dimensions~(following~\cite{hayes2018learning}), which samples from the standard normal distribution $\mathcal{N}(0,1)$. Subsequently, the UAP is scaled to control the distortion for the real data before being added on the input raw speech data with the real label \textit{Tom}. 
  The~\textit{Speaker Recognition} model, which is fixed and well-trained, is spoofed by the adversarial examples, which are the input speech data with UAPs, and predicts the input as~\textit{Jerry} by mistake.
  \\
  \indent Since the UAPs are not unique~\cite{Moosavi2017Universal}, we use a  ~\textit{Generator} to learn the mapping from the normal distribution into the UAP subspace. We use several \textit{UpBlock}s to synthesize the high-dimensional UAPs from the low-dimensional noise, and the convolution layer, batchnorm~\cite{ioffe2015batch}, and ReLU~\cite{nair2010rectified} are used in each \textit{UpBlock}.
  
  \subsection{Optimization}
  \label{sec:optimization}
  The optimization objective of our model is to find the adversarial examples with the smallest distortion, and they can attack the well-trained speaker recognition model successfully. Given the input noise $z$, the raw speech data $s$ and its class label $y$, the goal above can be formulated as follows:
  \begin{align}
    g = R_{x, \delta} - \lambda D_{x, x+\delta},
  \end{align}
  where $R$ denotes the attack success \textbf{R}ate, $D$ denotes the \textbf{D}istortion, $\lambda$ is the hyper-parameter to get a trade-off between $R$ and $D$, $\delta=G(z)$ is the UAP. In the optimization this objective function will be maximized.
  \\
  \indent In non-targeted attacks, attacking successfully means the victim model predicts by mistake, so we can optimize the attack success rate by reducing the prediction probability for the true class, and increasing the prediction probability for any wrong class. To spoof the victim model with minimal cost, we increase the probability for the class which is the top-1 class except for the true class. So $R$ can be formulated as follows:
  \begin{align}
    R_{x, \delta} = \left\{
      \begin{aligned}
      \max_{j\neq y}{p_j} - p_{y},&\quad \text{if}\quad R<T \\
      T,&\quad \text{else,}\\
      \end{aligned}
    \right.
    \label{eq:non-targeted}
  \end{align}
  where $p = f(x+\delta)$ is the output of pre-softmax layer (logit) with the adversarial example as input, $T$ is a threshold to stop the optimize for this sample.
  In targeted attacks, the attack is successful as long as the prediction class is the target class. Given the target class $t$, $R$ can be formulated as follows:
  \begin{align}
   R_{x, \delta, t} = \left\{
    \begin{aligned}
    p_t - \max_{j\neq t}{p_j},&\quad \text{if}\quad R<T \\
    T,&\quad \text{else.}\\
    \end{aligned}
  \right.
  \label{eq:targeted}
  \end{align}
  \\
  \indent The distortion for UAPs can be measured in two aspects: the objective quality and the perceptual quality. 
  We use Signal-Noise Ratio~(SNR) and the Perceptual Evaluation of Speech Quality (PESQ) score~\cite{rix2001perceptual} to evaluate the quality of the adversarial examples with perturbations in objective and perceptual, respectively. SNR is defined as:
  \begin{align}
    \text{SNR}(x,x') = 10\log_{10}\frac{\|x\|_2}{\|x-x'\|_2},
  \end{align}
  where $x'=x+\delta$ is the adversarial example with the perturbation $\delta$.
  PESQ, as an ITU-T recommendation standard~\cite{itu2000pesq}, is an integrated model to measure the distortion for the speech in telephony. It is a full-reference algorithm with range $[-0.5, 4.5]$ to measure the perceptual quality of the speech after a temporal alignment. It is worth mentioning that PESQ is not differentiable, so we only use it in the testing phase. In the training phase, $D$ is only the $\text{SNR}$ and we just optimize $\text{SNR}$ by minimizing the $L_2$ norm of the perturbations.
  \subsection{Inference}
  The inference is not intuitive because the input data are with variable lengths. In the training phase, we can clip the data into slices with a fixed length, but in the testing phase, we can not just drop the data beyond the UAPs. In our implementation, we use a simple but effective method~\textit{repeat+clip} to repeat the UAP until it is longer than the input sample and then clip it to make them two matched. In our experiments, we will conduct comparison experiment to study the influence of different UAP lengths.
  \section{Experimental Results}
  \label{sec:experiments}
  \subsection{Datasets and Metric}
  \textbf{Datasets:} Following~\cite{ravanelli2018speaker}, which proposed our victim model, we conduct the experiments on TIMIT~\cite{garofolo1993darpa}~(462 speakers totally) and LibriSpeech~\cite{panayotov2015librispeech}~(2484 speakers totally) datasets. The training/testing split follows the official implementation of~\cite{ravanelli2018speaker}, in which 2310/1386 samples are used for training/testing in TIMIT, and 14481/7452 samples are used for training/testing in LibriSpeech.
  \\
  \textbf{Metric:} We use the sentence error rate~(SER) to represent the attack success rate in the non-targeted attack, and the prediction target rate~(PTR) is used in targeted attacks because the attack is successful as long as the prediction is the target class in the targeted attack. The distortion is measured by SNR~(for objective quality) and PESQ~(for perceptual quality), as introduced in subsection~\ref{sec:optimization}. We use the official open-source implementation~\footnote{https://github.com/dennisguse/ITU-T\_pesq} for PESQ in our experiments.
  \subsection{Implementation Details}
  Our \textit{Generator} can only synthesize UAPs with the fixed length, so we randomly select a slice with a fixed length from the raw speech data in training phase. In our experiments, we synthesize UAPs for 200ms, which is 3200 dimensional because the data are with a sampling rate of 16000. We use the pretrained victim model which is released by the author of~\cite{ravanelli2018speaker}~\footnote{https://github.com/mravanelli/SincNet}. The hyper-parameter $\lambda$ will be finetuned in our experiments and the scale factor is fixed as $1$ because the perturbations are constrained on a small scale by the distortion item in our optimization objective. The threshold $T$ for non-targeted/targeted attack is set as 10/0 after being finetuned to get a good trade-off between $R$ and $D$.
  Besides, we initialize the biases and weights of the last convolution layer as zero to ensure that no perturbation is added on the signals at the beginning of the training\footnote{The code, data, and pretrained models will be released soon.}. 
  
  \subsection{Non-Targeted Attack}
  We conduct the non-targeted attack on TIMIT and LibriSpeech datasets to demonstrate the effectiveness of our proposed model. The results of these two datasets are illustrated in Table.~\ref{tab:non-targeted}. We can observe from the results that:
  \begin{itemize}
    \item For non-targeted attacks, the UAPs exist and our model manages to map the normal distribution into the UAPs subspace because our model can synthesize the UAPs which can attack the well-trained speaker recognition model with high success rate.
    \item On the TIMIT dataset, with $\lambda=1500$, the UAPs generated by our model can attack the well-trained speaker recognition model with an SNR of 49.87dB and a PESQ of 3.00, which means that the noise is noticeable but not intrusive.
    \item On the LibriSpeech dataset, with $\lambda=2500$, the UAPs generated by our model can attack the well-trained speaker recognition model with an SNR of 31.15dB and a PESQ of 2.33, which means that the noise is noticeable and a little intrusive. 
    \item On both TIMIT and LibriSpeech datasets, by tuning $\lambda$, we can control the trade-off between the attack success rate and the adversarial example quality.
  \end{itemize}
  It is worth mentioning that~\textit{the random perturbations in non-targeted attack can also achieve a high SER as long as the perturbations are intense enough}, so a high SER here cannot provide enough evidence that our model has learned some useful~\textit{universal patterns}. In Section~\ref{sec:ablation_study}, we will compare the UAPs generated by our model with the random perturbations to show that our model has learned the~\textit{universal patterns}.
  
  
  
  \begin{table}[t]
    \begin{center}
    \caption{Non-targeted attack on TIMIT/LibriSpeech dataset} 
    \label{tab:non-targeted}
      \begin{threeparttable}
    \begin{tabular}{c|c|c|c|c}
      \hline
      Dataset & $\lambda$ & SER($\%$)$\uparrow$ & SNR(dB)$\uparrow$ & PESQ$\uparrow$
      \\
      \hline
      \hline
      \multirow{6}{*}{TIMIT} & - & $1.52^\star$ & - & - \\
      \cline{2-5}
       & 500  & 97.5 & 44.13 & 2.13 \\
       & 1000 & 94.9 & 46.76 & 2.40 \\
       & 1500 & 97.0 & 49.87 & 3.00 \\
       & 2000 & 93.9 & 49.77 & 2.92 \\
       & 2500 & 92.4 & 50.49 & 2.79 \\
      \hline
      \hline
      \multirow{6}{*}{Libri} & - & $0.30^\star$ & - & -  \\
      \cline{2-5}
       & 1500 & 99.7 & 28.87 & 2.09  \\
       & 2000 & 97.0 & 30.62 & 2.22  \\
       & 2500 & 96.3 & 31.15 & 2.33  \\
       & 3000 & 93.7 & 32.72 & 2.45  \\
       & 3500 & 94.7 & 33.68 & 2.54  \\
      \hline
    \end{tabular}
    \begin{tablenotes}
      \item $^\star$ Error rate without attack on TIMIT/LibriSpeech.
    \end{tablenotes}
  \end{threeparttable}
    \end{center}
    \vspace{-5mm}
    \end{table}
  
  \subsection{Targeted Attack}
  In this subsection, we show our model's effectiveness on the targeted attack. In the targeted attack, we fix $\lambda$ as 3000/2000 for TIMIT/LibriSpeech dataset, respectively. 
  We randomly select 5 speakers from TIMIT/LibriSpeech dataset as the targets, and attack the victim model to misclassify any input sample as the target class. 
  The attack results are illustrated in Table.~\ref{tab:targeted}. Some conclusions can be drawn from the results:
  \begin{itemize}
    \item For the targeted attack, the UAPs exist and our model is successful to synthesize UAPs for the targeted attack on both TIMIT and LibriSpeech datasets.
    \item On the TIMIT dataset, we can achieve a PTR of 97.2$\%$ on average with an SNR of 48.53dB and a PESQ of 2.48, which means that the noise is noticeable, and a little intrusive, given the fact that the targeted attack is more challenging than the non-targeted attack.
    \item On the LibriSpeech dataset, we can achieve a PTR of 64.1$\%$ on average with an SNR 29.94dB and a PESQ of 2.11. This is not as good as that on the TIMIT dataset, because the speaker number in LibriSpeech~(2484) is much more than that in TIMIT dataset~(462).
  \end{itemize}
  Besides, \textit{the high success rate here can demonstrate that our model has learned the universal patterns for the targeted attack}. 
  Although random perturbations are able to achieve a high SER in non-targeted attack, they will fail to achieve a high PTR in targeted attack because they are random. Thus a high PTR in targeted attack can demonstrate that our model has learned the useful~\textit{universal patterns}.
  
  
  \begin{table}[t]
    \begin{center}
    \caption{Targeted attack on TIMIT/LibriSpeech dataset} 
    \label{tab:targeted}
    \begin{tabular}{c|c|c|c|c}
      \hline
      Dataset & Target& PTR($\%$)$\uparrow$ & SNR(dB)$\uparrow$ & PESQ$\uparrow$
      \\
      \hline
      \hline
      \multirow{5}{*}{TIMIT}
       & 0   & 99.1 & 48.09 & 2.49  \\
       & 100 & 98.4 & 48.86 & 2.41  \\
       & 200 & 98.0 & 48.55 & 2.52  \\
       & 300 & 93.9 & 48.93 & 2.42  \\
       & 400 & 96.6 & 48.20 & 2.55 \\
       \cline{2-5}
       & avg & 97.2 & 48.53 & 2.48 \\
      \hline
      \hline
      \multirow{5}{*}{Libri}
       & 0    & 64.5 & 29.73 & 2.10  \\
       & 500  & 40.8 & 30.73 & 2.14  \\
       & 1000 & 64.9 & 30.13 & 2.08  \\
       & 1500 & 54.9 & 29.75 & 2.07  \\
       & 2000 & 95.5 & 29.38 & 2.18  \\
       \cline{2-5}
       & avg & 64.1 & 29.94 & 2.11 \\
      \hline
    \end{tabular}
    \end{center}
    \vspace{-5mm}
    \end{table}
  
  \section{Ablation Study}
  \label{sec:ablation_study}
  \subsection{The Length of UAPs}
  \begin{figure}[t]
    \centering{\includegraphics[width=0.99\linewidth]{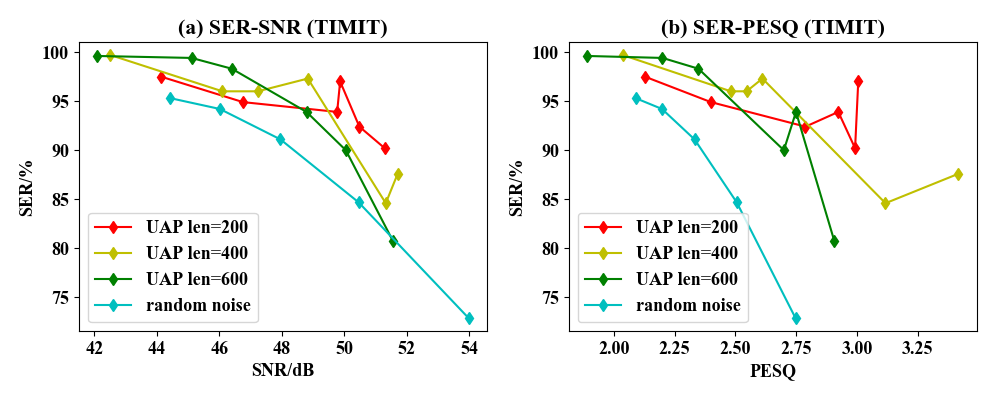}}
    \caption{The influence of the UAPs' length.}
    \label{fig:UAPs_length}
    \vspace{-5mm}
  \end{figure}
  We can only generate UAPs with a fixed length, but the input signals are with variable lengths. So we use~\textit{repeat+clip} method to make them two matched.
  The length of the UAPs may affect the performance of our model, so in this subsection, we conduct experiments to study how the UAP length influences the attacking performance. We generate the UAPs with duration 200ms, 400ms, and 600ms on TIMIT dataset, and we plot the SER-SNR and SER-PESQ curves to take both the adversarial examples quality and the attack success rate into account. Besides, we compare our generated UAPs with the~\textit{random noise}, which are the random perturbations sampled from the normal distribution $\mathcal{N}(0, \sigma^2)$~($\sigma$ is tuned to get five results with different trade-offs between SNR/PESQ with SER).
  \\
  \indent
  As illustrated in Fig.~\ref{fig:UAPs_length}, a curve is higher than another means that this model can achieve a higher SER with the same SNR/PESQ, indicating this model performs better than the other. From Fig.~\ref{fig:UAPs_length} (a), we can observe that:~(1)~the UAPs generated by our model perform better than the random perturbations, indicating that our model has learned the useful~\textit{universal patterns};~(2)~with SNR below 50dB, UAPs with different lengths achieve comparable performance, but with SNR higher 50dB, the shorter the UAPs are, the better performance they can achieve;~(3)~with UAPs length as 600ms, UAPs generated by our model may perform not as good as the random perturbations when SNR is higher than 52dB, the reason may be~\textit{the models for longer UAPs are more difficult to train but we train all models for different UAPs lengths for the same epochs to make a fair comparison.}
  From Fig.~\ref{fig:UAPs_length} (b), similar conclusions can be drawn except that the UAPs generated by our model performs much better with PESQ higher than 2.5 because the random perturbations struggle to achieve good perceptual quality~(PESQ).
  On both the objective quality~(SNR) and perceptual quality~(PESQ), the UAPs generated by our model perform better than the random perturbations, demonstrating that our model has learned the useful~\textit{universal patterns} to attack the well-trained speaker recognition model.

  \subsection{Noise Interpolation}
    To demonstrate our model is able to map the noise into the UAP subspace, we synthesize UAPs from the noise which is interpolated from two random noises and evaluate these UAPs on attacking the well-trained speaker recognition model. With an interpolation parameter $\beta$, the interpolation noise $z'$ can be obtained by:
    $
      z' = \beta z_1 + (1-\beta) z_2,
    $
    where $z_1$ and $z_2$ are two low-dimensional noise vectors sampled from the normal distribution $\mathcal{N}(0,1)$.
     As illustrated in Table~\ref{tab:noise_interpolation}, with different $\beta$, the UAPs generated by our model can achieve similar SER, SNR, and PESQ, validating our model can map $\mathcal{N}(0,1)$ into the UAPs subspace indirectly. 
  
  \begin{table}[t]
    \begin{center}
    \caption{Noise interpolation results~(SER) on TIMIT datasets} 
    \label{tab:noise_interpolation}
    \begin{threeparttable}
    \begin{tabular}{c|c|c|c|c|c|c}
      \hline
        $\beta$ & 0 & 0.2 & 0.4 & 0.6 & 0.8 & 1.0 
      \\\hline
        SER(\%)$\uparrow$ & 98.3 & 97.2 & 96.9 & 94.5 & 97.5 & 98.0  \\
        SNR(dB)$\uparrow$ & 49.6 & 50.0 & 50.1 & 50.5 & 49.8 & 49.5  \\
        PESQ$\uparrow$    & 2.99 & 3.02 & 3.04 & 3.05 & 3.01 & 2.99  \\
       \hline
    \end{tabular}
  \end{threeparttable}
    \end{center}
    \vspace{-5mm}
    \end{table}

  \section{Conclusion}
  \label{sec:conclusion}
  In this paper, we attempted to demonstrate the existence of the UAPs for the speaker recognition, and we proposed a generative network to map the low-dimensional noise space into the UAPs subspace to synthesize the UAPs efficiently. Experimental results showed that our model can generate UAPs and fool the state-of-the-art speaker recognition model with high success rate. The ablation study provided enough evidence to show that our model had learned useful \textit{universal patterns} for attacking the well-trained speaker recognition model. We envision our work to provide a benchmark for universal attacks for speaker recognition.
  
\bibliographystyle{IEEEbib}
\bibliography{icme2020template}

\end{document}